# Preparation, phase stability and characterization of $KMnO_4$-treated $Na_xCoO_2 \cdot yH_2O$

C.-J. Liu,[a*] C.-Y. Liao,[a] S. Neeleshwar,[b] and Y.-Y. Chen[b]

[a]*Department of Physics, National Changhua University of Education, Changhua 500, Taiwan, R. O. C.*

[b]*Institute of Physics, Academia Sinica, Taipei, Taiwan, R. O.C.*

## Abstract

We have successfully prepared superconducting phase of $Na_xCO_2 \cdot yH_2O$ with $T_c = 4.6$ K using $KMnO_4$ as the oxidizing agent. It is the first example to use oxidant other than $Br_2/CH_3CN$ solution to form the superconducting $Na_xCO_2 \cdot yH_2O$. According to ICP-AES analyses, it is found that the higher molar ratio of $KMnO_4$ relative to the Na content in the parent material leads to more deintercalation of Na from the parent material. At the molar ratio of $KMnO_4/Na \geq 4.286$, the XRD pattern of the product is similar to that of so-called "y = 0.6" phase of intermediate hydrate. In the talk, the phase stability and physical characterization of $KMnO_4$-treated $Na_xCO_2 \cdot yH_2O$ will be also presented.

## 1. Introduction

Takada et al. obtained the superconducting phase of $Na_xCoO_2 \cdot yH_2O$ with $T_c \approx 5$ K by immersing $Na_{0.7}CoO_2$ powders in $Br_2/CH_3CN$ solution followed by filtering and rinsing [1]. This process is generally considered as a chemical oxidation by removing Na partially before the $H_2O$ is intercalated between the $CoO_2$ layers and Na layers.

We have recently reported that superconducting $Na_xCoO_2 \cdot yH_2O$ phase can be obtained by immersing $Na_{0.7}CoO_2$ in $KMnO_4$ aqueous solution at low molar ratio of $KMnO_4$ relative to Na in the parent material [6]. The sodium content of the $KMnO_4$-treated $Na_xCoO_2 \cdot yH_2O$ can be varied by the relative molar ratio of $KMnO_4$. In this paper, we report a novel route to prepare the superconducting $Na_xCoO_2 \cdot yH_2O$ phase using $KMnO_4$ as an oxidizing agent. It is the first example to use oxidant rather than $Br_2/CH_3CN$ solution to form the superconducting phase of $Na_xCoO_2 \cdot yH_2O$.

## 2. Experimental

Polycrystalline parent materials of sodium cobalt oxides $Na_{0.7}CoO_2$ were prepared using a rapid heat-up procedure in order to avoid the loss of Na in the heating process [5]. The mixed powders of $Na_2CO_3$ and CoO were thoroughly ground using a Retch MM2000 laboratory mixer mill and calcined in a preheated box furnace at 700°C for 12 h. The resulting powders (0.5 -1 g) were immersed and stirred in 50 - 680 ml of $KMnO_4$ aqueous solution with different molar ratios of $KMnO_4$/Na labeled as 0.05X – 40X at room temperature for 5 days. The products were filtered and washed with deionized water several times. The powders were then stored in a chamber with relative humidity of 98% to avoid loss of the water content. Powder X-ray diffraction (XRD) patterns were obtained using a. Shimadzu XRD-6000 diffractometer equipped with Fe Kα radiation. The sodium content was determined by using a Perkin Elmer Optima 3000 DC inductively coupled plasma - atomic emission spectrometer (ICP-AES). Before the chemical analysis, samples are dehydrated by heating at 300℃ in air for 12 h. Thermogravimetric analysis (TGA) was carried out by using a Perkin Elmer Pyris 1 thermogravimetric analyzer. A commercial SQUID magnetometer (Quantum Design) was used to characterize the superconducting transition temperature of the samples.

## 3. Results and discussion

### 3.1 Phase formation of $KMnO_4$-treated $Na_xCoO_2 \cdot yH_2O$

In Fig. 1, we show the powder x-ray diffraction patterns of cobalt oxyhydrates obtained by treating $Na_{0.7}CoO_2$ using different molar ratios of $KMnO_4$/Na. It is clearly seen that different molar ratio of $KMnO_4$/Na (different X) results in variation of XRD patterns. The XRD patterns can be classified as three categories according to the molar ratio of $KMnO_4$/Na used to treat the parent material, which are $KMnO_4/Na \leq 0.1$, $0.3 \leq KMnO_4/Na \leq 2.29$, and $4.286 \leq KMnO_4/Na \leq 40$. It is known that both the parent material $Na_{0.7}CoO_2$ and the $Br_2/CH_3CN$-treated hydrates $Na_xCoO_2 \cdot yH_2O$ exhibit a hexagonal structure having a space group $P6_3/mmc$ (space group # 194). For the $Br_2/CH_3CN$-treated hydrates $Na_xCoO_2 \cdot yH_2O$, the water molecule is inserted between sodium and $CoO_2$ layers, resulting in the $c$-axis expansion from $c \approx 10.96$ Å to $c \approx 19.62$ Å in the unit cell with little change in the $a$-axis [1]. As shown in Fig. 1, the characteristic x-ray diffraction peak of the maximum intensity with the Miller index (002) shifts from $2\theta \approx 20.3°$ (d spacing $\approx 5.47$ Å) for the parent material irradiated by the Fe Kα radiation to 2θ ≈ 11.3° (d spacing ≈ 9.81 Å) for our samples between 0.3X and 2.29X, indicating that the *c*-axis expands from c ≈ 10.94 Å to c ≈ 19.6 Å in the unit cell, which is consistent with those for $Br_2/CH_3CN$-treated samples. These results confirm that we have successfully prepared cobalt oxyhydrates $Na_xCoO_2 \cdot yH_2O$ using $KMnO_4$ as the oxidant and de-intercalation agent without resort to the $Br_2/CH_3CN$ solution. For samples between 0.05X and 0.1X, the XRD patterns are a mixture of a fully-hydrated phase and a phase with peak positions close to those of the parent material and bromine-treated samples with substoichiometric or stoichiometric $Br_2/CH_3CN$ solutions [7]. It has been shown that the $Br_2/CH_3CN$ -treated hydrate phase is not stable at ambient conditions and tends to lose water becoming an intermediate hydrate phase with y ≈ 0.6 and *c* ≈ 13.8 Å [8]. Surprisingly, samples between 4.286X and 40X clearly shows different XRD patterns from low X samples and have the characteristic peak of (002) reflection occurring at 2θ ≈ 16° (d spacing ≈ 6.95 Å) and resulting in *c* ≈ 13.9 Å. Both the XRD pattern and the size of *c*-axis are very similar to the so-called $Br_2/CH_3CN$-treated intermediate hydrate phase with y ≈ 0.6. The lattice constants and the sodium contents of the samples are summarized in Table I. Chemical analyses show that the sodium contents in $Na_xCoO_2 \cdot yH_2O$ systematically decreases with increasing molar ratio of $KMnO_4$/Na. The values of x are 0.38, 0.33, 0.25 for 0.3X, 0.5X, and 2.29X samples, respectively. These results confirm that the role of $KMnO_4$ is acting as an oxidizing agent to partially de-intercalate the Na from the structure and hence oxidize the electronically active $CoO_2$ layers. The *c*-axis of the unit cell in $Na_xCoO_2 \cdot yH_2O$ tends to increase with increasing molar ratio of $KMnO_4$/Na from19.649 Å for the 0.3X sample to 19.709 Å for the 2.29X sample but with little changes in the *a*-axis.

**3.2 Phase stability of $KMnO_4$-treated $Na_xCoO_2 \cdot yH_2O$**

It is now well known that the as-prepared $Na_xCoO_2 \cdot yH_2O$ is not stable in the ambient conditions [8]. This situation would obviously devastate the interpretation of transport property measurements. The phase stability of $KMnO_4$-treated $Na_xCoO_2 \cdot yH_2O$ is therefore worth further investigation. Fig. 2 shows the XRD of the

The water content of 0.3X sample are checked and determined by heating the sample in flowing $O_2$ at the slowest rate of 0.1℃/min available to the Perkin Elmer Pyris 1 thermogravimetric ananyzer (TGA). Fig. 2 indicates a multi-stage loss of water with relatively unstable intermediates,[12] being consistent with the thermally unstable nature of the fully hydrated phase, .[5,13] The water content of fully hydrated phase is estimated to contain 1.45 and 1.55 $H_2O$ pre formula unit by taking the weight loss at 320 ℃ and 600℃, respectively, as the fully dehydrated phase (y = 0), assuming no oxygen deficiency in the sample for the present estimations.

Fig. 3 shows the zero field cooled and field cooled magnetization data of 0.3X sample measured in a field of 10 Oe. A superconducting transition is observed at 4.6 K. The mass magnetization at 1.8 K is $-1.28\times10^{-2}$ emu/g in the zero field cooling measurements, which is approximately 31 % of the theoretical value for perfect diamagnetism.

## 4. Conclusion

In summary, we have synthesized the superconductive cobalt oxyhydrates $Na_xCoO_2\cdot yH_2O$ using $KMnO_4$ as an oxidizing agent instead of using $Br_2/CH_3CN$ solutions. The role of $KMnO_4$ is to de-intercalate the Na from the structure and hence oxidize the Co ion based on the electron neutrality. The higher molar ratio of $KMnO_4$ relative to Na content used to treat the samples leads to more removal of Na. We are continuing to study the effects on the phase formation of $Na_xCoO_2 \cdot yH_2O$ using higher molar ratio of $KMnO_4$/Na and investigate the structure-property correlations for samples with different x of the Na content.

**Acknowledgments**

This work is supported by the National Science Council of ROC, grant No. NSC 92-2112-M-018-005.

[8]

Figure and Figure captions

Fig.1. Powder x-ray diffraction patterns of $KMnO_4$-treated $Na_xCoO_2 \cdot yH_2O$ obtained by immersing $Na_{0.7}CoO_2$ in different molar ratio of $KMnO_4$ aqueous solution with respect to the Na content in the parent compound. The 0.5X and 40X represent the molar ratios of $KMnO_4$/Na are 0.5 and 40.

Fig. 2. Thermogravimetric analysis of $Na_xCoO_2 \cdot yH_2O$ (0.3X) with a heating rate of 0.1℃/min in flowing oxygen. The water content is determined by assuming a complete loss of water at 320℃.

Fig. 3. Zero field cooled and field cooled d.c. magnetization for $Na_xCoO_2 \cdot yH_2O$ (0.3X) with $T_c = 4.6$ K.

**Table I.** Sodium content and lattice constants of $Na_xCoO_2 \cdot yH_2O$ prepared using $KMnO_4$ as oxidant

| Molar ratio of $KMnO_4$/Na | Sodium content x[a] | *a* axis[b] Å | *c* axis[b] Å |
|---|---|---|---|
| 0.3X | 0.38 | 2.8233(7) | 19.649(2) |
| 0.5X | 0.33 | 2.8246(1) | 19.663(2) |
| 2.29X | 0.25 | 2.8238(3) | 19.709(2) |

[a]The error in weight % of each element in ICP-AES analysis is ±3 %, which corresponds to an estimated error of ±0.02 per formula unit.

[b]Lattic constants are determined by least squares refinement using the XRD data between 2θ of 5° and 90° based on a hexagonal lattice with space group $P6_3/mmc$.